\documentclass{article}

\pdfoutput=1

\usepackage{arxiv}

\usepackage[utf8]{inputenc} % allow utf-8 input
\usepackage[T1]{fontenc}    % use 8-bit T1 fonts
\usepackage{hyperref}       % hyperlinks
\usepackage{url}            % simple URL typesetting
\usepackage{booktabs}       % professional-quality tables
\usepackage{amsfonts}       % blackboard math symbols
\usepackage{nicefrac}       % compact symbols for 1/2, etc.
\usepackage{microtype}      % microtypography
\usepackage{lipsum}		% Can be removed after putting your text content
\usepackage{graphicx}
\usepackage{natbib}
\usepackage{doi}

\usepackage{amsmath}
\usepackage{babel}
\usepackage[utf8]{inputenc}
\usepackage[T1]{fontenc}
\usepackage{hyperref}
\usepackage{graphicx}
\usepackage[ruled,vlined,linesnumbered]{algorithm2e}
\usepackage{bm}
\usepackage{color}
\usepackage{xcolor}
\usepackage{ulem}
\usepackage{commath}
\usepackage{bm}
\usepackage{graphicx}
\usepackage{epstopdf}
\usepackage{url}
\usepackage{booktabs}
\usepackage{multirow}
\usepackage{dblfloatfix}
\usepackage{subfigure}
\usepackage{caption}
\usepackage{soul}

\def\bX{{\mathbf X}}
\def\bx{{\mathbf x}}
\def\bY{{\mathbf Y}}
\def\by{{\mathbf y}}

\def\bz{\mathbf{z}}

\title{A stochastic metapopulation state-space approach to modeling and estimating Covid-19 spread}

\author{{Yukun Tan}\\
	Department of Electrical and Computer Engineering\\
	Texas A\&M University\\
	College Station, TX, 77843\\
	\texttt{yukuntan@tamu.edu} \\
	\And
	{Durward Cator III} \\
	Department of Electrical and Computer Engineering\\
	Texas A\&M University\\
	College Station, TX, 77843\\
	\texttt{tac464@tamu.edu} \\
	\And
	{Martial Ndeffo-Mbah} \\
	Veterinary Integrative Biosciences\\
	College of Veterinary Medicine and Biomedical Sciences\\
	Texas A\&M University\\
	College Station, TX, 77843\\
	\texttt{mndeffo@cvm.tamu.edu} \\
	\And
	{Ulisses Braga-Neto} \thanks{Corresponding author}\\
	Department of Electrical and Computer Engineering\\
	Texas A\&M University\\
	College Station, TX, 77843\\
	\texttt{ulisses@tamu.edu} \\
}

\renewcommand{\headeright}{}
\renewcommand{\undertitle}{}
\renewcommand{\shorttitle}{}
\hypersetup{
pdftitle={A template for the arxiv style},
pdfsubject={q-bio.NC, q-bio.QM},
pdfauthor={David S.~Hippocampus, Elias D.~Striatum},
pdfkeywords={First keyword, Second keyword, More},
}

\begin{document}
\maketitle

\begin{abstract}
Mathematical models are widely recognized as an important tool for analyzing and understanding the dynamics of infectious disease outbreaks, predict their future trends, and evaluate public health intervention measures for disease control and elimination. We propose a novel stochastic metapopulation state-space model for COVID-19 transmission, based on a discrete-time spatio-temporal susceptible/exposed/infected/recovered/deceased (SEIRD) model. The proposed framework allows the hidden SEIRD states and unknown transmission parameters to be estimated from noisy, incomplete time series of reported epidemiological data, by application of unscented Kalman filtering (UKF), maximum-likelihood adaptive filtering, and metaheuristic optimization. Experiments using both synthetic data and real data from the Fall 2020 Covid-19 wave in the state of Texas demonstrate the effectiveness of the proposed model.
\end{abstract}

% keywords can be removed

\section{Introduction}
Infectious disease outbreaks remain a major threat to global health. This is especially the case for highly pathogenic and transmissible diseases with pandemic potential. These global threats were recently exemplified by the 2009 swine flu outbreak and the ongoing COVID-19 pandemic caused by the novel Severe Acute Respiratory Syndrome coronavirus 2 (SARS-CoV-2). To effectively mitigate and control the spread of an epidemic, it is paramount for public health decision-making to be informed by an accurate understanding of the dynamics of disease transmission and the potential impact of intervention measures. 
%such as An epidemic is the rapid spread of infectious disease from one person to another within a short
%period of time, such as Severe Acute Respiratory Syndrome (SARS), Highly pathogenic avian in-
%fluenza (HPAI), Coronavirus disease (COVID 19). Especially the COVID 19 which is breaking
%out from January 2020, has taken lives of several thousands day by day. 
To this end, epidemic models have become an important tool to help better understand epidemic dynamics, predict its future trends, evaluate the effectiveness of intervention measures, such as lock-down or vaccination, and ultimately control the epidemic.

Mathematical epidemiology models can be broadly divided into two main types: compartmental models and agent-based models. An agent-based model is a very detailed stochastic model where the agent represent single individuals within a population \cite{degli2008mitigation, perez2009agent, hunter2018open, chang2020modelling, chao2020modeling, koo2020interventions, kretzschmar2020isolation, kerr2020covasim}. Such models are generally complex and computationally expensive. In compartmental models, the population is subdivided into epidemiological compartments with each compartmental generally representing a health or disease progression stage  \cite{balcan2010modeling, dukic2012tracking, osthus2017forecasting, sebastian2017state, keeling2020efficacy, sameni2020mathematical, godio2020seir, kobayashi2020predicting}. Thus compartmental models are computationally simple, scalable, and capable of describing the dynamics of the number of people in each compartment throughout the course of an epidemic. In this work, we propose a nonlinear state-space compartmental model.

Compartmental model date back to the early twentieth century, most notably to the work by \cite{kermack1927contribution}, whose susceptible/infected/recovered (SIR) model was used for modeling the plague (London 1665-1666, Bombay 1906) and cholera (London 1865) epidemics \cite{dukic2012tracking}. More specifically, it is a weighted directed graph representation of a dynamic system. The susceptible refers to those healthy people who are susceptible to the disease and may get infected; the infected refers to those under infection; the recovered refers to those who recovered from infection and will be temporarily or permanently immune to the disease. However, there are several drawbacks in the original SIR model: 

\begin{itemize}
    \item [1)] It is a deterministic model, meaning the model always performs the same for a given initial condition, which cannot explain the unknowable randomness in the observations \cite{sebastian2017state, hooker2011parameterizing,  kobayashi2020predicting};
    \item [2)] It is only a temporal model, which does not consider the spread in geographical regions, e.g. human interaction induced by modern transportation \cite{zhong2009simulation};
    \item [3)] It assumes that all the parameters are known, which is not realistic. Parameter estimation from noisy observations is needed for better understanding and forecasting epidemics~\cite{hooker2011parameterizing}.
\end{itemize}

Many variants SIR models have been proposed to address some of these issues, but accurate state and parameter estimation from partial and noisy observations remains an open problem. To` handle this issue, our proposed framework embeds the classical compartmental model within a nonlinear state-space model. The state model is a spatial-temporal stochastic dynamic model that allows hidden states in a given location to change over time and the disease dynamics in one location to affect neighbouring locations through human movements between locations. The proposed framework consists of a multinomial state model based on a variant of the SIR model --- the SEIRD model \cite{rapolu2020time, piccolomiini2020monitoring, korolev2021identification} --- and an observation model to allow the assimilation of publicly available data, including daily testing rate, daily test positivity rate, specificity and sensitivity of the tests. In addition, the model considers the differential testing rate between symptomatic patients and asymptomatic and healthy individuals. 
The proposed framework is employed to estimate i) the hidden epidemic state vector $X^t = (S^t, E^t, I^t, R^t, D^t)$ and ii) epidemiological parameters, such as infection rate and the average infectious period, using noisy incomplete time series epidemic data. Accurate estimation of the true epidemic state and parameters is paramount for designing and evaluating the effectiveness of control strategies. 
 
\section{Mathematical model}

\subsection{Classical SEIRD model}

The SEIRD model predicts the time evolution of epidemic. It models the dynamic interaction of people between five different compartments, namely, the susceptible (S), the exposed (E), the infected (I), the recovered (R) and the deceased (D). The classic continuous-time SEIRD model can be described by the following equations:
\begin{equation}
    \begin{aligned}
    \frac{dS(t)}{dt} &\,=\, -\lambda_S S(t)I(t)\\
    \frac{dE(t)}{dt} &\,=\, \lambda_S S(t)I(t) - \lambda_E E(t) \\
    \frac{dI(t)}{dt} &\,=\, \lambda_E E(t) - (\lambda_R + \lambda_D) I(t)\\
    \frac{dR(t)}{dt} &\,=\, \lambda_R I(t)\\
    \frac{dD(t)}{dt} &\,=\, \lambda_D I(t)\\
    \end{aligned}
\end{equation}
with $S(t) + E(t) + I(t) + R(t) + D(t) \,=\, 1$, where $S(t)$, $E(t)$, $I(t)$, $R(t)$, $D(t)$ are the fraction of the population (the size of which is assumed to be constant over the time interval of interest) at time $t$ that, respectively, $S$ is not yet infected with the disease; $E$ has been exposed to the virus but does not show symptoms yet; $I$ is infective after the virus incubation period; $R$ has been infected and then recovered; and $D$ is deceased due to the epidemic. 

The model is governed by the following parameters: 
\begin{itemize}
    \item $\lambda_S$ is the infection rate, which is the probability that an individual moves from the S to the E compartment. The nonlinear term $\lambda_S I(t)S(t)$ models the infection speed, which depends not only on the infection rate, but also on the fraction of susceptible and infected people at time $t$;
    \item $\lambda_E$ is the probability that an individual moves from the E to the I compartment. It can be understood as the inverse of the average incubation time;
    \item $\lambda_R$ is the recovery rate, which is the probability than an individual moves from the I to the R compartment. It can be understodd as the  inverse of the average recovery time;
    \item $\lambda_D$ is the mortality rate, which is the probability than an individual moves from the I to the D compartment.
\end{itemize}

\subsection{Covid-19 epidemic metapopulation state-space model}

Given the complexity and reality of epidemics, many different implementations of the classical SEIRD model have been proposed \cite{loli2020monitoring, korolev2021identification, tiwari2020mathematical, rapolu2020time}. Two clear drawbacks of the classical SEIRD model is the inability to model variation over a geographic area, and the fact that that the SEIRD values are assumed to be directly observable. In this paper, we address these drawbacks by means of a novel nonlinear metapolulation state-space framework, where the state process provides a spatial-temporal discrete-time stochastic model of the evolution of the number of individuals in the SEIRD compartments over several geographical regions, while the observation process models noisy time  series of  reported  epidemiological  data, considering the accuracy of tests and different testing rates for symptomatic and asymptomatic people.

\subsubsection{State model}

Consider a discrete-time state process $\{\bX_i^t; i = 1, \dots, G; t = 0, 1, \dots\}$, where $\bX_i^t = (S_i^t, E_i^t, I_i^t, R_i^t, D_i^t)$ is a state vector, containing the number of individuals in the SEIRD compartments in geographical area $i$ at time $t$, where the unit of time is typically day or week. At time $t$, the total number of individuals in region $i$ is $N_i = S_i^t + E_i^t + I_i^t + R_i^t + D_i^t$. We assume that this number stays constant over the time interval of interest (e.g., no significant amounts of migration is assumed to occur during the time interval of interest; plus , births and non-specific deaths are assumed to approximately balance out). This results in a good approximation provided that the interval of time in consideration consists of weeks or a few months, especially in the case of a full-blown outbreak when, the impact of migration is negligible. 
%In addition, in some important cases, such as the ongoing Covid-19 epidemics, infants have no significant impact \cite{XX}. 

Before proceeding, we recall the definition of the binomial and multinomial probability distributions. A random variable $Z$ has a binomial distribution with parameters $n\geq 1$ and $0<p<1$, denoted by $Z \sim {\rm Binomial}(n,p)$, if
\begin{equation}
    P(Z = k) \,=\, \frac{n!}{k!(n-k)!}\, p^k(1-p)^{n-k}, \quad k=0,\ldots,n\,.
\end{equation}
where $n! = n \times n-1 \times \cdots 1$, with $0! =1$. Variable $Z$ is the number of $k$ occurrences of an outcome among $n$ identical independent trials that can result in the outcome with probability $p$. The expected value and variance of this variable are $E[Z] = np$ and ${\rm Var}(Z) = np(1-p)$, respectively. More generally, we define a random vector $(Z_1,\ldots,Z_M)$ to have a multinomial distribution with parameters $n\geq 1$ and $p_1,\ldots,p_M\geq 0$ such that $\sum p_i \leq 1$, denoted by $(Z_1,\ldots,Z_M) \sim {\rm Multinomial}(n,p_1,\ldots,p_M)$, if
\begin{equation}
    P(Z_1 = k_1,\ldots,Z_M = k_M) \,=\, \frac{n!}{k_1!\cdots k_M!(n\!-\!\Sigma k_i)!}\, p_1^{k_1}\cdots p_M^{k_M}(1\!-\!\Sigma p_i)^{n-\Sigma k_i}, \quad k_1,\ldots,k_M \geq 0,\: \Sigma k_i \leq n\,.
\end{equation}
Variables $Z_1,\ldots,Z_M$ are the number of occurrences of each of $M$ outcomes over $n$ identical independent trials that can result in outcome $m$ with probability $p_m$. It is clear that $Z_m \sim {\rm Binomial}(n,p_m)$, for $m=1,\ldots,M$ (thus $E[Z_m] = np_m$ and ${\rm Var}(Z_m) = np_m(1-p_m)$). In addition, the case $M=1$ results in the binomial distribution.

In our model, the state vector $\bX_i^t = (S_i^t, E_i^t, I_i^t, R_i^t, D_i^t)$ is assumed to evolve according to the following nonlinear stochastic model:
\begin{equation}
\begin{aligned}
    S_i^{t+1} \,&=\, S_i^{t} \,-\, \Sigma_j \,N_{S_{i,j}}^{t} \\[0.5ex]
    E_i^{t+1} \,&=\, E_i^{t} \,+\, \Sigma_j \,N_{S_{i,j}}^{t} \,-\, N_{E_i}^{t}\\[0.5ex]
    I_i^{t+1} \,&=\, I_i^{t} \,+\, N_{E_i}^{t} \,-\, N_{R_i}^{t} \,-\, N_{D_i}^{t} \\[0.5ex]
    %R_i^t \,&=\, R_i^{t-1} \,+\, N_{I_i}^{t-1} \,-\, D_{R_i}^{t-1}
    R_i^{t+1} \,&=\, R_i^{t} \,+\, N_{R_i}^{t} \\[0.5ex]
    D_i^{t+1} \,&=\, D_i^{t} \,+\, N_{D_i}^{t}
\end{aligned}
\end{equation}
for $t=0,1,\ldots$, where $N^{t}_{S_{i,j}}$ is the number of susceptible individuals at time $t$ in region $i$ who will become exposed at time $t+1$ due to contact with an infective individual from region $j$, $N^{t}_{E_i}$ is the number of exposed people at time $t$ in region $i$ who will become infective at time $t+1$, while $N^{t}_{R_i}$ and $N^{t}_{D_i}$ are the numbers of infected people at time $t$ in region $i$ who will become recovered or deceased at time $t+1$, respectively. 

We assume that each susceptible individual in region $i$ becomes exposed due to contact with an infective individual from region $j$ at time $t$ independently with a probability $\lambda_{S}c_{ij}I_j^t/N_j$ (these parameters are explained below). Therefore, the distribution of the numbers of new infections over the regions is multinomial: 
\begin{equation}
    (N_{S_{i,1}}^t,\ldots,N_{S_{i,G}}^t) \,\sim\, {\rm Multinomial}\left(S_i^t, \lambda_{S}\frac{c_{i1}I_1^t}{N_1}, \dots, \lambda_{S}\frac{c_{iG} I_G^t}{N_G}\right). 
\end{equation}
The infection rate $\lambda_S > 0$ is disease-specific, but also affected by public policies, such as mask wearing and social distancing. We assume that these factors are constant over the geographical area (e.g., a single political unit, such as a state or country) and time interval in the study, so that $\lambda_S$ is the same for all regions. If $i=j$, then the contact is internal to region $i$, and $c_{ii} =1$. If $i \neq j$, then $0\leq c_{ij} \leq 1$ models the relative amount of transient interchange of individuals between regions $i$ and $j$, due to commuting, tourism, and so on. If regions $i$ and $j$ are far apart, or one of them is not economically important, then $c_{ij}$ is close to zero. Note that $c_{ij} = c_{ji}$. The specific values of $c_{ij}$ are selected in our study based on the {\it gravity model} \cite{zipf1946p, truscott2012evaluating, chen2021correlation}; Finally, the ratio $I_j^t/N_j$ indicates that if there are more infective individuals in region $j$, they will spread the disease with higher probability. Note that the expected value of the (normalized) $j$-th flux is $E[N_{S_{i,j}}^t/N_i] = \lambda_{S}c_{ij}(S_i^t/N_i)(I_j^t/N_j)$. This may be contrasted to the flux $\lambda_{S}S(t)I(t)$ in the classical case. 

In our model, an exposed individual in region $i$ becomes infective at time $t$, independently of the other regions, with probability $\lambda_E$. Hence,
\begin{equation}
    N_{E_i}^{t} \,\sim\, {\rm Binomial}(E_i^t, \lambda_E)\,.
\end{equation}
The expected number of exposed individuals who become infective at time $t$ is thus $E[N_{E_i}^{t}] = \lambda_E E_i^t$. Note also that the distribution of the total time until an exposed individual becomes infective is {\it geometric} with parameter $\lambda_E$, so that the expected number of time units until an exposed individual becomes infective is $\lambda_E^{-1}$.

Finally, each infective individual in region $i$ becomes recovered or deceased at time $t$, independently of the other regions, with probabilities $\lambda_R$ and $\lambda_D$, respectively. Hence,  
\begin{equation}
    (N_{R_i}^{t}, N_{D_i}^{t}) \,\sim\, {\rm Multinomial}(I_i^t, \lambda_R, \lambda_D)\,.
\end{equation}
The expected numbers of infective individuals who become recovered or deceased at time $t$ are $E[N_{R_i}^{t}] = \lambda_R I_i^t$ and $E[N_{D_i}^{t}] = \lambda_D I_i^t$, respectively. The expected numbers of time units until an infective individual becomes recovered or deceased are $\lambda_R^{-1}$ and $\lambda_D^{-1}$, respectively.

\subsubsection{Observation model}
\label{Sec-ObsMod}

The observation model is a major contribution of this work. We model reported epidemiological data, namely confirmed new cases and cumulative recorded deaths, as noisy observations on the true state of the pandemic, as defined in the previous section. The proposed observation model addresses the uncertainty introduced by imperfect testing and the imbalance in testing of symptomatic and asymptomatic individuals (the former cohort is tested more~\cite{allen2020population}).

% , false positive and false negative rates of tests are taking into account. To be more precise, we split the total number of confirmed cases into two parts, one is the uninfected people but misdiagnosed as infected because of the false positive of the tests and the other one is the true infected people (the infected here are specifically refers to those suffering from COVID 19). Then in each part, symptomatic and asymptomatic aspects are considered. As we know, the uninfected people can also have a certain chance of having similar symptoms, e.g. influenza. Also, there are many infected people are asymptomatic. No matter infected or not, people with symptoms are more likely to be tested. In this procedure, $\varepsilon_1$ and $\varepsilon_2$ are correlated with $\varepsilon_3$ and $\varepsilon_4$ via two rates we can observe, namely, test positivity rate and testing rate. 

We model the reported data as a time series $\{\bY_i^t; i=1,\dots,G; t=0,1,\dots\}$ where $\bY_i^t = (P_i^t, Q_i^t)$ contains the numbers $P_i^t$ and $Q_i^t$ of new confirmed cases and deaths, respectively, in geographical area $i$ at time $t$. The number of reported new cases contain both false and true positives, 
% \begin{equation}
%     \bY_i^t \,=\, \mathbf{h}(\bX_i^t) + \mathbf{v_i^t}
% \label{eq:observation_model}
% \end{equation}
% for $t = 0, 1, \dots$ represents the time, and $i = 0, 1, \dots$ denotes the geographical region, where $\mathbf{h}$ is a general dynamics function which is a way to express the observations by the state variables  and $\mathbf{v}$ is the observation noise which is dependent on the state in our model.
\begin{equation}
    P_i^t \,=\, N_{TP_i}^t \,+\, N_{FP_i}^t,
\label{eq:seird_observation}
\end{equation}
for $t=0,1,,\ldots$, where $N_{TP_i}^t$ and $N_{FP_i}^t$ are the number of false and true testing positives, respectively. Let $\alpha$ and $\beta$ be the false positive and true positive rates, respectively, of the Covid-19 test under consideration (multiple tests of different accuracy can be introduced by splitting the populations according to the test received). True positives come from the exposed and infective populations. In addition, a percentage of the infective population is asymptomatic; we  assume that those are tested at a smaller rate than the symptomatic ones. Accordingly, we split the number of new true positives at time $t$ as the sum of the number of positive-tested symptomatic infective people and the number of positive-tested asymptomatic infective and exposed individuals: 
\begin{equation}
  N_{TP_i}^t \,\sim\, {\rm Binomial}(\varepsilon^t_2(1 - \varepsilon^t_4) I_i^t,\, \beta) \,+\, {\rm Binomial}(\varepsilon^t_1 (\varepsilon^t_4 I_i^t + E_i^t),\, \beta)\,,
\end{equation}
where $\varepsilon^t_1$ and $\varepsilon^t_2$ are the testing rates of asymptomatic and symptomatic individuals at time $t$, respectively, while $\varepsilon^t_4$ is the percentage of infective individuals who are asymptomatic at time $t$. These parameters are assumed to be the same over all regions. As shown later, these parameters can be calculated from the overall testing and positivity rates, which are available from public records.

Similarly, the false positives come from the susceptible and recovered populations, and here we have to distinguish between symptomatic individuals (who are not infected by Covid-19 but instead by other similar viruses, such as Influenza) and asymptomatic individuals:
\begin{equation}
    N_{FP_i}^t \,\sim\, {\rm Binomial}(\varepsilon^t_1(1-\varepsilon^t_3)(S_i^t + R_i^t),\, \alpha) \,+\, {\rm Binomial}(\varepsilon^t_2\varepsilon^t_3(S_i^t + R_i^t),\, \alpha)\,,
\end{equation}
where $\varepsilon^t_3$ is the percentage of non-specific symptomatic individuals at time $t$.

Finally, we assume that the reported cumulative recorded deaths in each region is a fraction of the true number $D_i^t$, according to 
\begin{equation}
   Q_i^t \,\sim\, {\rm Binomial}(D_i^t, \, \beta)\,.
\label{eq-obsD}
\end{equation}
Even though there might be a number of non-Covid-19 deaths falsely reported as Covid-19, we are assuming that this number is small and can be ignored. Notice that (\ref{eq-obsD}) implies that Covid-19 deaths are underreported, which is widely assumed to be the case. 

Let $R_T^t$ be the testing rate, i.e., the percentage of the population that is tested at time $t$, and let $R_P^t$ the positivity rate at time $t$. Both these parameters are reported by health authorities. Next, we show how the parameters $\varepsilon^t_1$, $\varepsilon^t_2$, $\varepsilon^t_3$, and $\varepsilon^t_4$ can be calculated from $R_T^t$ and $R_P^t$

The test positivity rate ($R_P^t$) is the percentage of positive tests $P^t$ over the total number of tests $N_{T}^t$ in a given region; it can be written as
\begin{equation}
    \begin{aligned}
    R_P^t \,=\, \frac{P^t}{N_T^t} \,=\, \frac{\varepsilon^t_1 A \alpha +   \varepsilon^t_2 B \alpha + \varepsilon^t_2 C \beta + \varepsilon^t_1 D \beta}{\varepsilon^t_1 A + \varepsilon^t_2 B + \varepsilon^t_2 C + \varepsilon^t_1 D}
    \end{aligned}
\end{equation}
where $A = (1 - \varepsilon^t_3)(S^t + R^t)$, $B = \varepsilon^t_3 (S^t + R^t)$, $C = (1 - \varepsilon^t_4) I^t$, and $D = \varepsilon^t_4 I^t + E^t$.

Similarly, the testing rate $R_T^t$ is the percentage of total tests over the total population, which can be written as
\begin{equation}
    R_T^t \,=\, \frac{N_T^t}{N^t} \,=\, \frac{\varepsilon^t_1 A + \varepsilon^t_2 B + \varepsilon^t_2 C + \varepsilon^t_1 D}{A + B + C + D}
\end{equation}

%\noindent $\varepsilon^t_{flu}$ is the fraction of the population that is tested on a given day;\\

%\noindent , since we don't consider the general death in the short period, we don't consider the false positive part of death;\\

%\noindent The observation noise is the same style as the state model.\\

Using the previous two equations, we can solve for $\varepsilon^t_1$ and $\varepsilon^t_2$ in terms of $\varepsilon^t_3$ and $\varepsilon^t_4$:
% \begin{equation}
%     \begin{aligned}
%         &{\varepsilon^t_1}^t \,=\, \frac{(S^t + E^t + I^t + R^t){R_T}^t}{1 - \beta - \alpha} \times \\[0.5ex]
%         & \left[\frac{{R_P}^t(\varepsilon^t_3 (S^t + R^t)+(1 - \varepsilon^t_4) I^t)}{\varepsilon^t_3 (S^t + R^t)(\varepsilon^t_4 I^t + E^t) - (1 - \varepsilon^t_3)(S^t + R^t)(1 - \varepsilon^t_4) I^t}\right.\\[0.5ex]
%         & - \left.\frac{\alpha \varepsilon^t_3 (S^t + R^t) + (1 - \beta)(1 - \varepsilon^t_4) I^t}{\varepsilon^t_3 (S^t + R^t)(\varepsilon^t_4 I^t + E^t) - (1 - \varepsilon^t_3)(S^t + R^t)(1 - \varepsilon^t_4) I^t} \right]\\[0.5ex]
%         &{\varepsilon^t_2}^t \,=\, \frac{(S^t + E^t + I^t + R^t){R_T}^t}{1 - \beta - \alpha} \times \\[0.5ex]
%         & \left[\frac{\alpha(1 - \varepsilon^t_3)(S^t + R^t) + (1 - \beta)(\varepsilon^t_4 I^t + E^t)}{\varepsilon^t_3 (S^t + R^t)(\varepsilon^t_4 I^t + E^t) - (1 - \varepsilon^t_3)(S^t + R^t)(1 - \varepsilon^t_4) I^t} \right.\\[0.5ex]
%         & - \left. \frac{{R_P}^t((1 - \varepsilon^t_3)(S^t + R^t)+(\varepsilon^t_4 I^t + E^t))}{\varepsilon^t_3 (S^t + R^t)(\varepsilon^t_4 I^t + E^t) - (1 - \varepsilon^t_3)(S^t + R^t)(1 - \varepsilon^t_4) I^t} \right]\\[0.5ex]
%         \end{aligned}
%         \label{eq:epsilon_1}
% \end{equation}
\begin{equation}
    \begin{aligned}
        {\varepsilon^t_1} \,=\,& \frac{(S^t + E^t + I^t + R^t)R_T^t({R_P}^t(\varepsilon^t_3 (S^t + R^t)+(1 - \varepsilon^t_4) I^t) - \alpha \varepsilon^t_3 (S^t + R^t) - \beta(1 - \varepsilon^t_4) I^t)}{(\beta - \alpha)(\varepsilon^t_3 (S^t + R^t)(\varepsilon^t_4 I^t + E^t) - (1 - \varepsilon^t_3)(S^t + R^t)(1 - \varepsilon^t_4) I^t)}\\[0.5ex]
        {\varepsilon^t_2} \,=\,& \frac{(S^t + E^t + I^t + R^t)R_T^t}{\beta - \alpha} \times \\[0.5ex]
        & \frac{\alpha(1 - \varepsilon^t_3)(S^t + R^t) + \beta(\varepsilon^t_4 I^t + E^t) - {R_P}^t((1 - \varepsilon^t_3)(S^t + R^t)+(\varepsilon^t_4 I^t + E^t))}{\varepsilon^t_3 (S^t + R^t)(\varepsilon^t_4 I^t + E^t) - (1 - \varepsilon^t_3)(S^t + R^t)(1 - \varepsilon^t_4) I^t}\\[0.5ex]
        \end{aligned}
        \label{eq:epsilon_1}
\end{equation}

Now, $\varepsilon^t_1$ and $\varepsilon^t_2$ are the testing rates of asymptomatic and symptomatic individuals at time $t$, and can be related to each other. According to \cite{allen2020population}, we have
\begin{equation}
    \varepsilon^t_2 = \zeta \varepsilon^t_1,
    \label{eq:epsilon_2}
\end{equation}
where $\zeta$ is uniformly distributed in the interval $[2, 4.3]$ to account for 
various non-specific symptoms (e.g., fever, cough, loss of taste/smell) that are shared by Covid-19 and other common viral diseases.
Substituting (\ref{eq:epsilon_2}) into (\ref{eq:epsilon_1}) results, after some algebra, in the following expression for ${\varepsilon^t_3}$
% \begin{equation}
%     \begin{aligned}        
%         &{\varepsilon^t_3}^t \,=\, \frac{\zeta(1 - \beta - {R_P}^t)(\varepsilon^t_4 I^t + E^t)}{(1 - \zeta)(S^t + R^t)({R_P}^t - \alpha)} \\[0.5ex]
%         & + \frac{(1 - \varepsilon^t_4) I^t (1 - \beta - {R_P}^t)}{(1 - \zeta)(S^t + R^t)({R_P}^t - \alpha)}\\[0.5ex]
%         & - \frac{\zeta(S^t + R^t)({R_P}^t - \alpha)}{(1 - \zeta)(S^t + R^t)({R_P}^t - \alpha)}\\[0.5ex]
%     \end{aligned}
% \end{equation}
\begin{equation}
        {\varepsilon^t_3} \,=\, \frac{(\beta - R_P^t)(\varepsilon^t_4 I^t + E^t) + \zeta(1 - \varepsilon^t_4) I^t (\beta - R_P^t) - (S^t + R^t)(R_P^t - \alpha)}{(\zeta - 1)(S^t + R^t)(R_P^t - \alpha)}
\end{equation}
In addition, according to \cite{buitrago2020occurrence}, the percentage of asymptomatic infective people $\varepsilon^t_4$ is around 0.2 and from this the other parameters $\varepsilon^t_1$, $\varepsilon^t_2$, $\varepsilon^t_3$ can be calculated.

\section{Covid-19 epidemic state and parameter estimation}

In this section, we describe state-of-the-art estimators for the state and parameters of the metapopulation state-space model introduced in the previous section. These estimators use an input the noisy time series $\bY_i^t = (P_i^t,Q_i^t)$ of reported new cases and deaths in the several geographical regions in the study. The state vector $\bX_i^t = (S_i^t, E_i^t, I_i^t, R_i^t, D_i^t)$ is estimated at each time $t$ by using the Unscented Kalman Filter (UKF) \cite{wan2000unscented}, which is the state-of-the-art estimator for stochastic nonlinear state-space models, while the parameters $\lambda_S, \lambda_R, \lambda_D$ are estimated by maximum-likelihood, by using the metaheuristic optimization - fish school search algorithm \cite{bastos2008novel, bastos2013enhanced, tan2020pallas}.

First, we write the state-space model in the previous section in the standard form
\begin{equation}
\begin{aligned}
     \bX_i^{t+1} &\,=\, f\,(\bX_i^t) + \mathbf{n}_i^t\\[0.5ex]
     \bY_i^t & \,=\, h\,(\bX_i^t) + \mathbf{v}_i^t
 \label{eq:SS_model}
\end{aligned}
\end{equation}
where ${f}$ and ${h}$ are nonlinear mappings, and $\{\mathbf{n}_i^t; i = 1, \dots, G; t = 0, 1, \dots\}$ and
$\{\mathbf{v}_i^t; i = 1, \dots, G; t = 0, 1, \dots\}$ are
white-noise (i.e., uncorrelated in time) transition and observation noise processes, respectively, which are independent of each other and independent of the initial state $\{\bX_i^0; i=1,\ldots,G\}$.

The proposed state-space model can be put in the standard form (\ref{eq:SS_model}) by rewriting the state equations as:
\begin{equation}
\begin{aligned}
    S_i^{t+1} \,=\,& \left(1 \,-\, \lambda_S\,\Sigma_j\, c_{ij} I_j^{t}/N_j\,\right) \times S_i^{t} \,+\, n_{S_i}^{t} \\[1ex]
    E_i^{t+1} \,=\,& (1\,-\,\lambda_E) \times E_i^{t} \,+\, \left(\lambda_S\,\Sigma_j\, c_{ij} I_j^{t}/N_j\right) \times S_i^{t} \,+\, n_{E_i}^{t} \\[1ex]
    I_i^{t+1} \,=\,& (1 \,-\, \lambda_R \,-\, \lambda_D) \times I_i^{t} \,+\, \lambda_E \times E_i^{t} \,+\, n_{I_i}^{t} \\[1ex]
    R_i^{t+1} \,=\,&  R_i^{t} \,+\, \lambda_R \times I_i^{t} \,+\, n_{R_i}^{t} \\[1ex]
    D_i^{t+1} \,=\,&  D_i^{t} \,+\, \lambda_D \times I_i^{t} \,+\, n_{D_i}^{t}
\label{eq:seird_state}
\end{aligned}
\end{equation}
where the transition noise terms are:
%$n_i^t \,=\, [n_{S_i}^t, n_{E_i}^t, n_{I_i}^t, n_{R_i}^t, n_{D_i}^t]$, 
\begin{equation}
\begin{aligned}
    n_{S_i}^t & \,=\, -\Sigma_j \left(N^t_{S_{i,j}}-E[N^t_{S_{i,j}}]\right)\\
    & \,=\, -\Sigma_j\left(N^t_{S_{i,j}} - \lambda_S\,c_{ij}S_i^{t}I_j^{t}/N_j\right) \\[1ex]
    n_{E_i}^t &\,=\, -\left(N_{E_i}^t - E[N_{E_i}^t]\right) \,+\, \Sigma_j \left(N^t_{S_{i,j}}-E[N^t_{S_{i,j}}]\right)\\
    &\,=\, -\left(N_{E_i}^t-\lambda_E E_i^t\right) \,+\, \Sigma_j\left(N^t_{S_{i,j}} - \lambda_S\,c_{ij}S_i^{t}I_j^{t}/N_j\right)\\[1ex]
    n_{I_i}^t & \,=\, \left(N_{E_i}^t - E[N_{E_i}^t]\right) \,-\, \left(N_{R_i}^t - E[N_{R_i}^t]\right) \,-\, \left(N_{D_i}^t - E[N_{D_i}^t]\right) \\
    &\,=\, \left(N_{E_i}^t - \lambda_E E_i^t\right) \,-\, \left(N_{R_i}^t - \lambda_R I_i^t\right) \,-\, \left(N_{D_i}^t - \lambda_D I_i^t\right) \\[1ex]
    n_{R_i}^t &\,=\, \left(N_{R_i}^t - E[N_{R_i}^t]\right) \,=\, \left(N_{R_i}^t - \lambda_R I_i^t\right) \\[1ex]
    n_{D_i}^t &\,=\, \left(N_{D_i}^t - E[N_{D_i}^t]\right) \,=\, \left(N_{D_i}^t - \lambda_D I_i^t\right) \\[0.5ex]
\end{aligned}
\end{equation}

Similarly, the observation model can be rewritten in the standard form (\ref{eq:SS_model}):
\begin{equation}
    \begin{aligned}
        P_i^t &\,=\, \beta\,\varepsilon^t_2\left(1 - \varepsilon^t_4\right) I_i^t \,+\, \beta\,\varepsilon^t_1\left(\varepsilon^t_4 I_i^t + E_i^t\right) \\
         &\quad + \alpha\, \varepsilon^t_1\left(1 - \varepsilon^t_3\right)\left(S_i^t + R_i^t\right)  \,+\, \alpha\,\varepsilon^t_2 \varepsilon^t_3\left(S_i^t + R_i^t\right) \,+\, v_{P_i}^t\\[1ex]
        Q_i^t &\,=\, \beta\, D_i^t \,+\, v_{Q_i}^t\\
    \end{aligned}
\end{equation}
where the observation noise terms are:
\begin{equation}
    \begin{aligned}
         v_{P_i}^t &\,=\, N_{TP_i}^t - E[N_{TP_i}^t] \,+\, N_{FP_i}^t - E[N_{FP_i}^t]\\
         &\,=\, N_{TP_i}^t - \beta\, (\varepsilon^t_2(1 - \varepsilon^t_4) I_i^t \,+\, \varepsilon^t_1(\varepsilon^t_4 I_i^t + E_i^t)) \\
        & \quad+\, N_{FP_i}^t - \alpha\,( \varepsilon^t_1(1 - \varepsilon^t_3)(S_i^t + R_i^t)  \,+\, \varepsilon^t_2 \varepsilon^t_3(S_i^t + R_i^t))\\[1ex]
        v_{Q_i}^t &\,=\, D_{TP_i}^t - E[D_{TP_i}^t] \,=\, D_{TP_i}^t - \beta\,D_i^t
    \end{aligned}
\end{equation}

\subsection{Unscented Kalman filter}

With the state-space model in the standard format (\ref{eq:SS_model}), one can apply the {\it Unscented Kalman Filter} (UKF) algorithm \cite{wan2000unscented, simon2006optimal, sarkka2013bayesian} to estimate the state variables $(S_i^t, E_i^t, I_i^t, R_i^t, D_i^t)$ from the noisy data $(P_i^t,Q_i^t)$ of new cases and cumulative deaths in geographical area $i$ at time $t$. The UKF assumes that the statistics of the state variables at time $t=0$ are known (in practice, these values need to be only very roughly guessed since, as the time $t$ increases, the UKF generally ``forgets'' the information in the initial state). It also assumes that all the parameters are known; however, in the next section we describe a methodology to estimate the parameters from the data as well.

% time series of data. $\bX_i^t; i = 1, \dots, G; t = 0, 1, \dots\}$ from the observed data   

% $\{\bX_i^k; i=1,\ldots,G\}$.application of the Uncented Kalman Filter

% With the nonlinear mappings 
% The UKF is the filter that produce several sigma points around the current state with its covariance. Then, propagate these points by using the nonlinear map to get more accurate mean and covariance. Based on , the procedure we used is shown below (every region is calculated parallel in the same procedure):

% Recall our state space model eq~(\ref{eq:state_model}) and eq~(\ref{eq:observation_model}), the state variables belong to:
% \begin{equation}
%     \bX \sim N(\mathbf{m}, \mathbf{P})
% \end{equation}
% and the process noise and observation noise belong to:
% \begin{equation}
%     \begin{aligned}
%     \mathbf{n} &\sim N(0, \mathbf{Q})\\
%     \mathbf{v} &\sim N(0, \mathbf{R})
%     \end{aligned}
% \end{equation}

%Let Given the state $\bX_{t|t} = (S_t, E_t, I_t, R_t, D_t)$ be the state vector 
In the UKF algorithm, $\mathbf{m}_{t|t}$ and $\mathbf{P}_{t|t}$ denote the estimates at time $t$ of the mean and error covariance matrix, respectively, of the state vector $\mathbf{X}^t = (S^t, E^t, I^t, R,D^t)$ using all observed data up to time~$t$ (for brevity, the subscript $i$ used previously to discriminate the region is omitted throughout this section, since the same process is applied to all regions separately). We initalize the mean vector via\linebreak $E[I^0] = E[E^0] = P^0$, $E[R^0] = 0$, $E[D^0] = Q^0/\beta$, and $E[S^0] = N-E[I^0+E^0+R^0+D^0]$,  while the covariance matrix is initialized to the identity matrix. The UKF estimate of the state $\mathbf{X}^t$ is $\mathbf{m}_{t|t}$, with uncertainty given by~$\mathbf{P}_{t|t}$. These estimates are computed by the following iteration:

\vspace{3mm}
{\bf Initialization:} Given the initial observation $\mathbf{Y}^0 = (P^0,Q^0)$, let
\begin{equation}
  \mathbf{m}_{0|0} \,=\, (N-2P^0-D^0/\beta,\,P^0,\,P^0,\,0,\,Q^0/\beta)\,, \quad \mathbf{P}_{0|0} = {\rm I}\,.
\end{equation}

\vspace{3mm}
For $t = 1,2,\ldots$ {\bf repeat}:
\vspace{3mm}

{\bf Prediction:}

1) Generate (2n+1) sigma points, 
\begin{equation}
\begin{aligned}
\bz_{t-1|t-1}^{0} &\,=\, \mathbf m_{t-1|t-1}\,, \\[0.5ex]
\bz_{t-1|t-1}^{i} &\,=\, \mathbf m_{t-1|t-1} + \left[ \sqrt{n\mathbf{P}_{t-1|t-1}}\right]_i, \quad i \,=\, 1, \dots, n\,, \\[0.5ex]
\bz_{t-1|t-1}^{i+n} &\,=\, \mathbf m_{t-1|t-1} - \left[ \sqrt{n\mathbf{P}_{t-1|t-1}}\right]_{i-n}, \quad i \,=\, n+1, \dots, 2n\,, \\[0.5ex]
\end{aligned}
\end{equation}

where $\left[ \sqrt{n\mathbf{P}_{t-1|t-1}}\right]_i$ is the i-th  column  of  the  matrix  square  root.

2) Propagate the sigma points through the state equation:
\begin{equation}
\begin{aligned}
\bx_t^i &\,=\, \mathbf{f}(\bz_{t-1|t-1}^i)\,, \quad\, i \,=\, 0, \dots, 2n \\[0.5ex]
\end{aligned}
\end{equation}

3) Compute predicted mean and predicted error covariance:
\begin{equation}
\begin{aligned}
\mathbf{m}_{t|t-1} &\,=\, \frac{1}{2n}\sum_{i=0}^{2n} \, \hat{\bx}_t^i\,,\\[0.5ex]
\mathbf{P}_{t|t-1} &\,=\, \frac{1}{2n}\sum_{i=0}^{2n}(\hat{\bx}_t^i - \mathbf{m}_{t|t-1}) \times (\hat{\bx}_t^i - \mathbf{m}_{t|t-1})^T + \mathbf{Q}_{t-1}\,,\\[0.5ex]
\end{aligned}
\end{equation}
\hspace{5ex}where $\mathbf{Q}_{t-1}$ is the covariance matrix of the transition noise (see the Appendix for its derivation).

\vspace{3mm}

{\bf Update:}

1) Update sigma points based on the predicted mean and error covariance:
\begin{equation}
\begin{aligned}
\bz_{t|t-1}^0 \,=\,& \mathbf m_{t|t-1}\,, \\[0.5ex]
\bz_{t|t-1}^i \,=\,& \mathbf m_{t|t-1} + \left[ \sqrt{n\mathbf{P}_{t|t-1}}\right]_i, \quad i \,=\, 1, \dots, n\,, \\[0.5ex]
\bz_{t|t-1}^{i+n} \,=\,& \mathbf m_{t|t-1} - \left[ \sqrt{n\mathbf{P}_{t|t-1}}\right]_{i-n},\quad i \,=\, n+1, \dots, 2n\,. \\[0.5ex]
\end{aligned}
\end{equation}

2) Propagate the sigma point through the observation equation:
\begin{equation}
\begin{aligned}
\hat{\by}_t^i &\,=\, \mathbf{h}(\bz_{t|t-1}^i)\,, \quad i \,=\, 0,\dots, n\,.
\end{aligned}
\end{equation}

3) Compute predicted measurement mean, measurement covariance matrix, and  cross-covariance matrix:
\begin{equation}
\begin{aligned}
\bm{\mu}_t &\,=\, \frac{1}{2n}\sum_{i=0}^{2n} \hat{\by}_t^i\,,\\[0.5ex]
\mathbf{S}_t &\,=\, \frac{1}{2n}\sum_{i=0}^{2n} (\hat{\by}_t^i - \bm{\mu}_t)(\hat{\by}_t^i - \bm{\mu}_t)^T + \mathbf{R}_t\,,\\[0.5ex]
\mathbf{C}_t &\,=\, \frac{1}{2n}\sum_{i=0}^{2n}(\bz_{t|t-1}^i - \mathbf{m}_{t|t-1})(\hat{\by}_t^i - \bm{\mu}_t)^T,\\[0.5ex]
\end{aligned}
\end{equation}
\hspace{5ex}where $\mathbf{R}_{t-1}$ is the covariance matrix of the observation noise (see the Appendix for its derivation).

4) Compute the filter gain and new state error covariance matrix. Assimilate the observation $\mathbf{Y}^t$ at time $t$ to find new state mean vector $\mathbf{m}_{t|t}$:
\begin{equation}
\begin{aligned}
\mathbf{K}_t &\,=\, \mathbf{C}_t \mathbf{S}_t^{-1}, \\[0.5ex]
\mathbf{m}_{t|t} &\,=\, \mathbf{m}_{t|t-1} + \mathbf{K}_t(\mathbf{Y}^t - \bm{\mu}_t), \\[0.5ex]
\mathbf{P}_{t|t} &\,=\, \mathbf{P}_{t|t-1} - \mathbf{K}_t \mathbf{S}_t \mathbf{K}_t^T.\\
%\mathbf{v}_t &\,=\, \hat{\bY}_t - \bm{\mu}_t.
\end{aligned}
\end{equation}

\subsection{Maximum-likelihood adaptive filtering}

The UKF algorithm requires that all parameter values be known. The parameters that govern the proposed model in (\ref{eq:seird_state}) and (\ref{eq:seird_observation}) are $\lambda_S$, $\lambda_E$, $\lambda_R$, $\lambda_D$, $\alpha$, $\beta$, $\varepsilon_1$, $\varepsilon_2$, $\varepsilon_3$, and $\varepsilon_4$. Of these, $\alpha$ and $\beta$ are known, while $\varepsilon_1$ through $\varepsilon_4$ can be obtained using the procedure described in Section~\ref{Sec-ObsMod}. Among the $\lambda$ parameters, some may be known a priori, but others may be unknown. Let $\bm{\theta}$ be a vector containing these parameters. Accurate estimation of $\bm{\theta}$ from the observed data is key to make the proposed methodology useful in practice. We estimate $\bm{\theta}$ using maximum-likelihood method in combination with the UKF, which is known as maximum-likelihood adaptive filtering \cite{ito2000gaussian, wu2006numerical, kokkala2015sigma}. 
%All of the variables following are only consider one region, so the corner mark of the region is ignored.

Let $\bY^{0:t} = \{\bY^0,\bY^1,\ldots,\bY^t\}$ denote the observed data up to time $t$. The log-likelihood of the model parameters $\bm{\theta}$ at time~$t$ is given by:
\begin{equation}
\begin{aligned}
L_t(\bm{\theta}) &\,=\, \log \ p_{\bm{\theta}}(\bY^{0:t}) \,=\, \log \left[\,p_{\bm{\theta}}(\bY_t \mid \bY^{0:t-1})p_{\bm{\theta}}(\bY^{t-1} \mid
\bY^{0:t-2}) \cdots p_{\bm{\theta}}(\bY^1\mid\bY^0)p_{\bm{\theta}}(\bY^0)\right]\\
&\,=\, L_{t-1}(\bm{\theta}) + \log\, p_{\bm{\theta}}(\bY^t \mid \bY^{0:t-1}) \,,
\label{eq:likelihood}
\end{aligned}
\end{equation}
where
\begin{equation}
\begin{aligned}
p_{\bm{\theta}}(\bY^t \mid \bY^{0:t-1}) \,=\,& - \frac{1}{2} \,\log \,|2\pi\,\mathbf{S}_t(\bm{\theta})| \,-\, \frac{1}{2} \,\mathbf{V}_t^T(\bm{\theta})\, \mathbf{S}_t^{-1}(\bm{\theta})\,\mathbf{V}_t(\bm{\theta})\,,
\end{aligned}
\end{equation}
with $\mathbf{V}_t \,=\, \bY^t - \bm{\mu}_t$. The quantities $\bm{\mu}_t$ and $\mathbf{S}_t$ are calculated in the UKF recursion. Then, the target is to maximize the log-likelihood $L_t(\theta)$,
\begin{equation}
    \hat{\bm{\theta}}_{ML} \,=\, \arg\max_{\bm{\theta}} \ L_t(\bm{\theta})\,.
\end{equation}
There are several possible ways to address this optimization problem. For example, one can apply the Expectation-Maximization (EM) algorithm, which is especially effective when there is a closed-form solution for the ``M'' maximization step, which avoids recursive gradient calculation. However, there is no such closed-form solution in our case. Other gradient-based optmization methods did not produce good results. Instead, we used 
biology-inspired metaheuristic optimization, in this case, the {\it Fish School Search} (FSS) algorithm \cite{bastos2008novel, bastos2013enhanced}, which we had already applied with success in our previous work~\cite{tan2020pallas}. For the details about the FSS algorithm, please see \cite{bastos2008novel, bastos2013enhanced, tan2020pallas}.

\section{Numerical experiments}
\label{sec:exp}

In this section, we present the results of numerical experiments, using synthetic and real data from the state of Texas in the United States. We divided Texas into 11 regions, based on the Texas Health and Human Services (HHS) regional map.

Sections \ref{sec:exp1} and \ref{sec:exp3} assume the parameter values displayed in Table~\ref{tab:table1} (unless stated otherwise). The COVID-19 test false positive rate $\alpha$ and false negative rate $1 - \beta$ are based on \cite{asai2020covid}; the infection rate $\lambda_S$, mean incubation period $1/\lambda_E$, and mean recovery time $1/\lambda_R$ are consistent with CDC data; while the mortality rate $\lambda_D$ is based on publicly reported data in Texas. The values for  $\varepsilon^t_1$ through $\varepsilon^t_4$ are computed as described in Section~\ref{Sec-ObsMod}. On the other hand, in Section \ref{sec:exp4}, the methodology was applied to real Covid-19 epidemic data.
% (e.g.reported cases, reported deaths, test positivity rate, testing rate), save for the COVID-19 test false positive rate, false negative rate, and $\lambda_E$, which are treated as known data (a reasonable assumption). 

\begin{table}[h]
\caption{Parameter values for numerical experiments (section 4.1, 4.2)} %title of the table
\centering % centering table
\begin{tabular}{c@{\hskip 1in}c} %creating eight columns
\hline\hline %inserting double-line
\bf{Parameter} & \bf{Value} \\
\hline\hline % inserts single-line
Test false positive rate ($\alpha$) & 0.01\\ % Entering row contents
\hline
Test false negative rate ($1-\beta$) & 0.15\\
\hline
Infection rate ($\lambda_S$) & 0.4\\ % [1ex] adds vertical space
\hline % inserts single-line
Mean incubation period ($1/\lambda_E$) & 10 (days)\\
\hline
Mean recovery time ($1/\lambda_R$) & 14 (days)\\
\hline
Mortality rate ($\lambda_D$) & 0.01\\
\hline
$\varepsilon^t_1, \varepsilon^t_2, \varepsilon^t_3, \varepsilon^t_4$ & 0.03,  0.1, 0.1, 0.2\\
\hline
\end{tabular}
\label{tab:table1}
\end{table}

\begin{figure}[htbp]
\centering
\includegraphics[width=160mm]{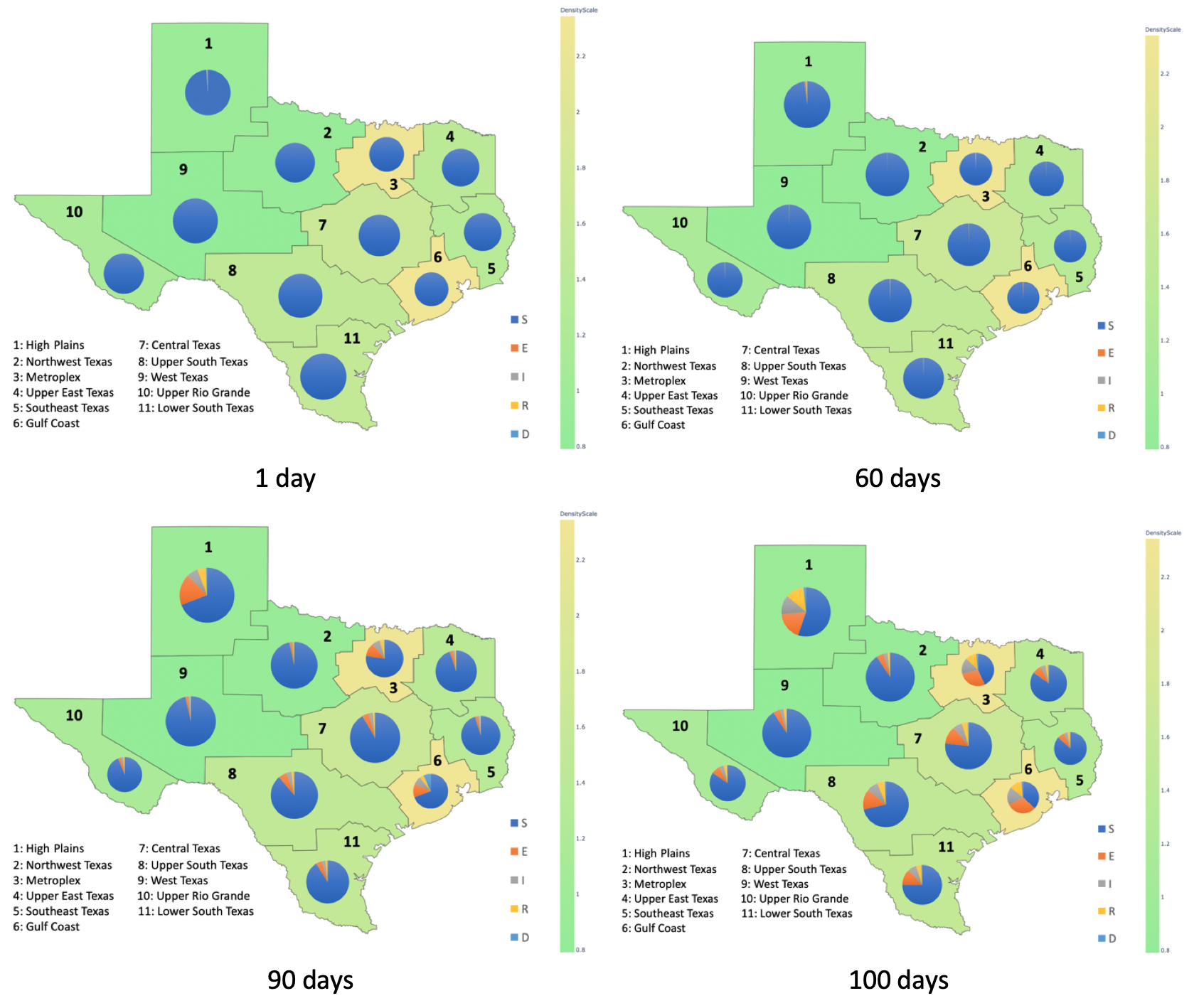}
\caption{Prediction of Covid-19 spread over the state of Texas after 1, 60, 90, and 100 days from initial infection.}
\label{fig:geo}
\end{figure}

\subsection{Forward prediction of epidemic dynamics}
\label{sec:exp1}

%In the following cases of simulation, we would like to show our 
Here we illustrate the proposed model's ability to predict the spatial and temporal dynamics of epidemic, assuming the parameter values displayed in Table~1. We ran the model for a period of 200 days. Fig~\ref{fig:geo} displays the results at day 1, day 60, day 90, and day 100. In this simulation, the epidemic is assumed to have originated in region 1 with only a few infective individuals, as can be seen in the upper left diagram of Fig~\ref{fig:geo}. After 60 days, the pandemic has spread to other regions, but it is still very limited. However, after 90 days, i.e., only 30 days after the previous snapshot, the epidemic has spread much more widely, especially in the originating region 1, as well as regions with large populational density, e.g. region 3 and region 6, which is where Dallas and Houston are located, respectively. Finally, after only another 10 days, the number of infected individuals has almost doubled. These results show that the epidemic may be easier to control at an early stage. They also show that if the epidemic is allowed to run its course, it will spread at an exponential rate after a period of time. These predictions underscore the need for early-stage public health interventions, such as social distancing,  mask wearing, or a limited lockdown in the critical originating region 1.

\subsection{State and parameter estimation from synthetic data}
\label{sec:exp3}

In this section, we investigate the ability of the proposed maximum-likelihood adaptive filtering methodology in recovering both the hidden state and all unknown parameters of the pandemic from a synthetic time series of reported new cases and deaths. The synthetic data allow us to evaluate the performance of the methodology against the simulated ground truth.

\begin{figure}[htbp]
\centering
\includegraphics[width=140mm]{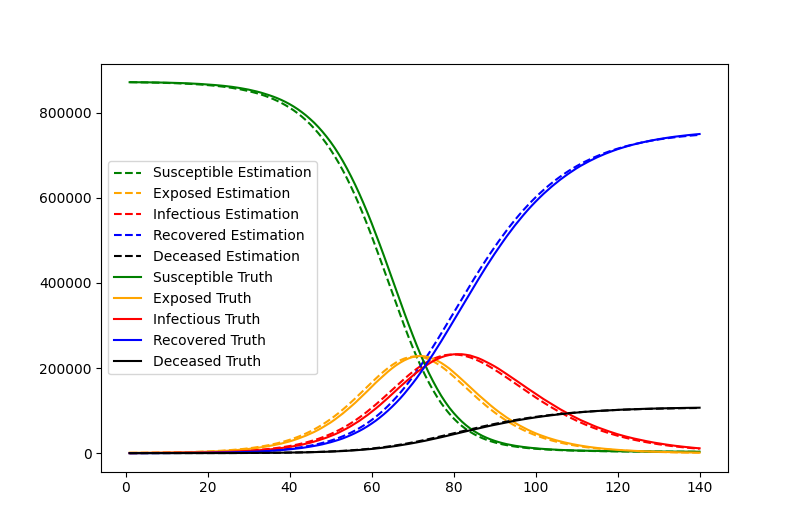}
\caption{State estimation in region 1 with all parameter values known.}
\label{fig:known}
\end{figure}

In the first experiment, we assume that the parameter values (see Table~\ref{tab:table1}) are known and we evaluate the ability of the filtering methodology to recover the hidden SEIRD state from the synthetic times series. Although in practice not all parameters would be known, this experiment allows us to evaluate the pure state estimation capabilities of the algorithm. 
Figure~\ref{fig:known} displays the results over region 1 (the behavior was similar over all other regions). We can see that the maximum-likelihood adaptive filter can track the state evolution remarkably well. 
% Figure~\ref{fig:known_relative_error} displays the  relative error, defined as (estimated state-true state)/(true state), for each of the state variables. We can see that the error is small, and becomes smaller as more data is accumulated.

% \begin{figure}[htbp]
% \centering
% \includegraphics[width=120mm]{figs/relative_error.png}
% \caption{Relative error in region 1 with all parameter values known.}
% \label{fig:known_relative_error}
% \end{figure}

\begin{figure}[htbp]
\centering
\includegraphics[width=140mm]{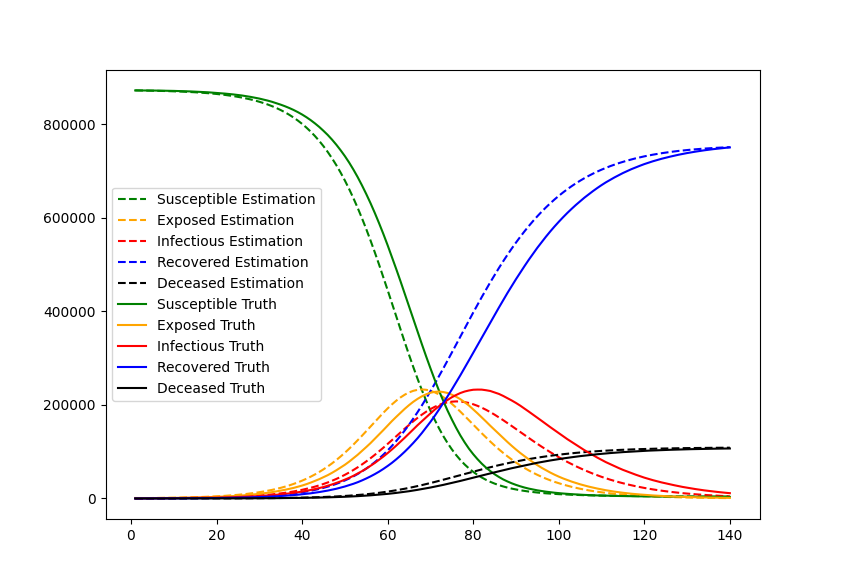}
\caption{State estimation in region 1 using estimated parameter values.}
\label{fig:unknown}
\end{figure}

In the second experiment, we remove the assumption that the infection rate, mean recovery time, and mortality rate parameters are known, and evaluate the performance of the methodology in recovering the values of these parameters. This is a difficult problem, since the states are also unknown, and the algorithm must perform simultaneous state and parameter estimation. Table~\ref{tab:table2} shows that the algorithm produced estimated parameter values that are close to the groundtruth values. Figure~\ref{fig:unknown} displays state estimation results when the estimated parameters are plugged in for the unknown parameters. We can see that the results are not quite as good as the case where all parameters are known, in Figure~\ref{fig:known}, but the proposed methodology is still able to track the state evolution well. 
% However, we can see that the error is still remarkably small, as can be seen in Figure~\ref{fig:unknown_relative_error}, which shows that the relative error for all states is in most cases under $0.5\%$, after enough observation data have been accumulated. 

\begin{table}[h]
\caption{Parameter estimation with synthetic data.} %title of the table
\centering % centering table
%\begin{tabular}{c c c}
\begin{tabular}{c@{\hskip 0.4in}c@{\hskip 0.4in}c} %creating eight columns
\hline\hline %inserting double-line
\bf{Parameter}& \bf{Groundtruth} & \bf{Estimated Value} \\
\hline\hline % inserts single-line
Infection rate ($\lambda_S$) & 0.4 & 0.451\\ % [1ex] adds vertical space
\hline % inserts single-line
Mean recovery time ($1/\lambda_R$) & 14 (days) & 12.05 (days)\\
\hline
Mortality rate ($\lambda_D$) & 0.01 & 0.0121\\
\hline
\end{tabular}
\label{tab:table2}
\end{table}

% \begin{figure}[htbp]
% \centering
% \includegraphics[width=120mm]{figs/estimate_relative_error.png}
% \caption{Relative error in region 1 using estimated parameter values.}
% \label{fig:unknown_relative_error}
% \end{figure}

\subsection{State and Parameter Estimation using Johns Hopkins University Covid-19 Data}
\label{sec:exp4}

\begin{figure}[h!]
\centering
\includegraphics[width=140mm]{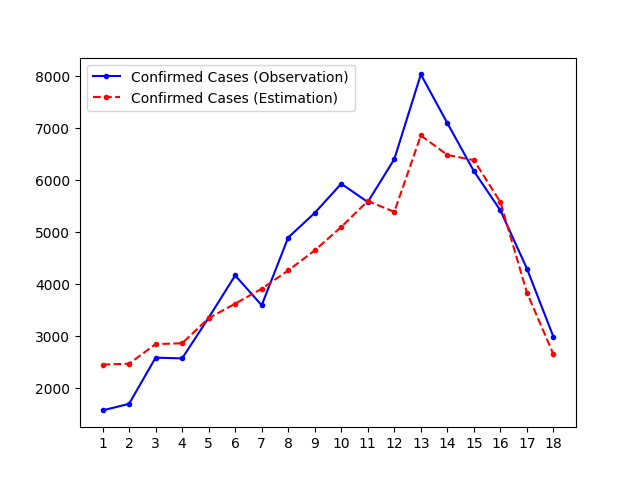}
\caption{Prediction of new cases using JHU Covid-19 data over region 3. The blue line represents the actual observed data, the red dashed line are the predicted values in the time interval where the observed data is known.}
\label{fig:confirmed}
\end{figure}

In this experiment, we demonstrate the performance of the algorithm on Texas Covid-19 data from the Center for Systems Science and Engineering (CSSE) at Johns Hopkins University \cite{jhu}. Due to the presence of delayed reporting, there is often incorrect data in the early part of the week and corrections at the end of the week. To address this, we consider the time unit to be week, not day, and average the daily reported data (from Sunday to Saturday) to produce one data point for the week. The data used in this experiment is from the week of Oct 4th, 2020 to the week of Feb 1st, 2021, which is from the third (Fall 2020) wave of the COVID-19 epidemic in the United States. 
% The first 16 time points (weeks) (from Oct 4th, 2020 to Jan 18th, 2021) were used as training data to estimate all unknown parameters, and the remaining  2 time points (weeks) were used to evaluate the ability of the algorithm to predict the future state of the epidemic. 
In this experiment, we only use data from four big regions (3, 6, 7, 8), which contain Dallas, Houston, Austin and San Antonio, which are the most reliable data available. 

\begin{figure}[h!]
\centering
\includegraphics[width=140mm]{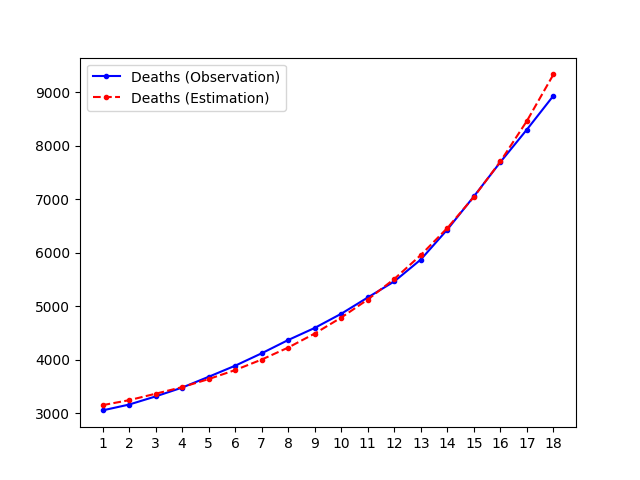}
\caption{Prediction of cumulative deaths using JHU Covid-19 data over region 3. The blue line represents the actual observed data, the red dashed line are the predicted values in the time interval where the observed data is known.}
\label{fig:death}
\end{figure}

We assume that only the test false positive rate, false negative rate and the mean incubation time are known. However, the value of the latter in Table~\ref{tab:table1} is based on a daily time unit, so here the value used is 10/7, or $\lambda_E = 0.7$. All estimated rates obtained by the algorithm in this experiment can be divided by 7 in order to obtain daily rates.

\begin{table}[h]
\caption{Estimated parameters for Texas Fall 2020 wave using JHU Covid-19 data} %title of the table
\centering % centering table
%\begin{tabular}{c c c}
\begin{tabular}{c@{\hskip 0.3in}cccc} %creating eight columns
\hline\hline %inserting double-line
%Parameter&\multicolumn{4}{c}{Region 3}{Region 6}{Region 7}{Region 8} \\ [0.5ex]
\bf{Parameter} & \bf{Region 3} & \bf{Region 6} & \bf{Region 7} & \bf{Region 8}\\
\hline\hline % inserts single-line
Infection rate ($\lambda_S$) & 0.16 & 0.16 & 0.16 & 0.16\\ % [1ex] adds vertical space
\hline % inserts single-line
Mean recovery time ($1/\lambda_R$) & 9.92 (days) & 10.52 (days) & 10.85 (days) & 10.95 (days)\\
\hline
Mortality rate ($\lambda_D$) & 0.0111 & 0.00992 & 0.0135 & 0.0112\\
\hline
\end{tabular}
\label{tab:table3}
\end{table}

In addition, unlike in the experiment with synthetic data, we have no groundtruth for the states or parameters. In order to evaluate the performance of the algorithm, we use the estimated values of states and parameters to predict the time series of observed data and check its agreement with the actual JHU data. These plots are displayed in Figures \ref{fig:confirmed} and \ref{fig:death} over region 3. They indicate that our model can accurately estimate the hidden states and unknown parameters. Table~\ref{tab:table3} displays the estimated values of the parameters. The value for the infection rate is the same over all regions, by assumption; however, Figures \ref{fig:confirmed} and \ref{fig:death} provide evidence that this is a reasonable assumption. 

\subsection{Discussion}

Compared with the standard SEIRD model, which assumes that the states of the epidemic can be observed directly, our  stochastic state-space model realistically treats these states as hidden and only indirectly observed from the reported data. The proposed methodology is also able to incorporate detailed information about the testing procedure, such as false positive and false negative test rates and differential testing rate between symptomatic and asymptomatic patients. Our model provides a more accurate estimate of unknown/unobserved epidemiological parameters, such as the time-varying reproductive number, which are paramount for monitoring disease transmission dynamics and evaluating the effectiveness of ongoing control strategies. Moreover, our state-space model is more suitable for designing robust optimal control strategies and providing more accurate predictions of the future trends of an epidemic, based on the noisy reported time series data, than standard epidemic models.

Our model has limitations, which can be addressed by future work. First, our model does not consider vital dynamics such as birth and death from other causes. Provided that the model is run over a short interval of time, such as an isolated wave in the epidemic, the results will not be affected too much; in addition, new born play an insignificant role in Covid-19 spread. Second, we assume that individuals infected with COVID-19 will develop life-long immunity. Though some epidemiological data have indicated evidence of COVID-19 reinfections, in most cases immunity persist for at least eight months. Therefore, we believe that this assumption has a marginal impact on our results, provided once again that the model is run over a short enough time interval. Third, our model ignored the impact of age on COVID-19 epidemiological parameters such as infection rate, disease severity (symptomatic vs asymptomatic), mortality rate, recovery rate, and testing rate. Age-specific infection rate and disease-induced complications have been observed during the ongoing COVID-19 pandemic and considered to play an import role in disease transmission and the design of public health policies. Fourth, our model in it current form assumes that connection between the different regions can be modeled with a simple gravity commuting model \cite{zipf1946p, truscott2012evaluating, chen2021correlation}. This limitation can be addressed by using empirical population mobility data. Finally, Our model did not consider potential temporal changes of parameters over time, e.g. the mortality rate may be be high and recovery rate low at the beginning of an outbreak because of lack of effective treatments, or public behavior, such as social distancing and mask wearing, may vary over time. Also, in the case of the Fall 2020 wave, it is known that mass vaccination began at the end of this wave (December 2020), which very likely had a major impact in the disease transmission parameters. Hence, the estimated parameter values reflect average rates over the time interval in consideration. 

\section{Conclusion}

We proposed a novel stochastic metapopulation state-space model for COVID-19 transmission, based on a discrete-time SEIRD model, which is able to estimate the hidden epidemic states and transmission parameters from a noisy time series of reported epidemiological data, by applying unscented Kalman filtering (UKF), Maximum Likelihood (ML) adaptive filtering, and metaheuristic optimization. We reported results from a comprehensive set of experiments, using synthetic data and real epidemic data to demonstrate that our model can estimate parameters and predict future trends accurately and effectively. The proposed framework was applied to the state of Texas in the United States, using data from the Center for Systems Science and Engineering (CSSE) at Johns Hopkins University. It can also be applied to any city/state/nation by providing the necessary time series data (e.g. reported cases, deaths, and testing rates). The proposed methodology may provide a valuable tool for revealing the current epidemic status through accurate estimated states and disease transmission parameters, improving the ability of authorities to make informed decisions on public health measures to contain the disease. 

%\begin{itemize}
 %   \item We have not considered the number of birth and natural deaths since we believe the population is balanced in the short period without the epidemic and also it is uncommon that the virus will transport from mothers to new birth.
 %   \item We assume the epidemic disease will be immunized after the infected which means the person in the recovery ($R$) category will not be reintroduced to susceptible. 
 %   \item The infection rates among different ages are not considered, although we know that higher aged people are more vulnerable.
 %   \item The model is suitable to apply in some fixed period meaning that it is not accurate to fit the model to the epidemic from day one till the end because although we model the stochastic property of the parameters, we have not considered the trend of parameters over time, e.g. death rate will be high and recovery rate will be low at the beginning because of the lack of the effective medicine and experience. Also the different policy over time will also affect the parameters, e.g. ask people to keep the social distance and wear mask will greatly reduce the infection rate.
%\end{itemize}

\section*{Conflict of interest}

\noindent
The authors declare that they have no conflicts of interest.

\section*{Acknowledgements}

\noindent
Martial Ndeffo-Mbah acknowledges funding from the National Science Foundation RAPID Award [grant number DEB-2028632].

\bibliographystyle{unsrtnat}
\bibliography{template}

\begin{thebibliography}{40}
\providecommand{\natexlab}[1]{#1}
\providecommand{\url}[1]{\texttt{#1}}
\expandafter\ifx\csname urlstyle\endcsname\relax
  \providecommand{\doi}[1]{doi: #1}\else
  \providecommand{\doi}{doi: \begingroup \urlstyle{rm}\Url}\fi

\bibitem[Degli~Atti et~al.(2008)Degli~Atti, Merler, Rizzo, Ajelli, Massari,
  Manfredi, Furlanello, Tomba, and Iannelli]{degli2008mitigation}
Marta Luisa~Ciofi Degli~Atti, Stefano Merler, Caterina Rizzo, Marco Ajelli,
  Marco Massari, Piero Manfredi, Cesare Furlanello, Gianpaolo~Scalia Tomba, and
  Mimmo Iannelli.
\newblock Mitigation measures for pandemic influenza in italy: an individual
  based model considering different scenarios.
\newblock \emph{PloS one}, 3\penalty0 (3):\penalty0 e1790, 2008.

\bibitem[Perez and Dragicevic(2009)]{perez2009agent}
Liliana Perez and Suzana Dragicevic.
\newblock An agent-based approach for modeling dynamics of contagious disease
  spread.
\newblock \emph{International journal of health geographics}, 8\penalty0
  (1):\penalty0 1--17, 2009.

\bibitem[Hunter et~al.(2018)Hunter, Mac~Namee, and Kelleher]{hunter2018open}
Elizabeth Hunter, Brian Mac~Namee, and John Kelleher.
\newblock An open-data-driven agent-based model to simulate infectious disease
  outbreaks.
\newblock \emph{PloS one}, 13\penalty0 (12):\penalty0 e0208775, 2018.

\bibitem[Chang et~al.(2020)Chang, Harding, Zachreson, Cliff, and
  Prokopenko]{chang2020modelling}
Sheryl~L Chang, Nathan Harding, Cameron Zachreson, Oliver~M Cliff, and Mikhail
  Prokopenko.
\newblock Modelling transmission and control of the covid-19 pandemic in
  australia.
\newblock \emph{Nature communications}, 11\penalty0 (1):\penalty0 1--13, 2020.

\bibitem[Chao et~al.(2020)Chao, Oron, Srikrishna, and
  Famulare]{chao2020modeling}
Dennis~L Chao, Assaf~P Oron, Devabhaktuni Srikrishna, and Michael Famulare.
\newblock Modeling layered non-pharmaceutical interventions against sars-cov-2
  in the united states with corvid.
\newblock \emph{medRxiv}, 2020.

\bibitem[Koo et~al.(2020)Koo, Cook, Park, Sun, Sun, Lim, Tam, and
  Dickens]{koo2020interventions}
Joel~R Koo, Alex~R Cook, Minah Park, Yinxiaohe Sun, Haoyang Sun, Jue~Tao Lim,
  Clarence Tam, and Borame~L Dickens.
\newblock Interventions to mitigate early spread of sars-cov-2 in singapore: a
  modelling study.
\newblock \emph{The Lancet Infectious Diseases}, 20\penalty0 (6):\penalty0
  678--688, 2020.

\bibitem[Kretzschmar et~al.(2020)Kretzschmar, Rozhnova, and van
  Boven]{kretzschmar2020isolation}
Mirjam Kretzschmar, Ganna Rozhnova, and Michiel van Boven.
\newblock Isolation and contact tracing can tip the scale to containment of
  covid-19 in populations with social distancing.
\newblock \emph{Available at SSRN 3562458}, 2020.

\bibitem[Kerr et~al.(2020)Kerr, Stuart, Mistry, Abeysuriya, Hart, Rosenfeld,
  Selvaraj, Nunez, Hagedorn, George, et~al.]{kerr2020covasim}
Cliff~C Kerr, Robyn~M Stuart, Dina Mistry, Romesh~G Abeysuriya, Gregory Hart,
  Katherine Rosenfeld, Prashanth Selvaraj, Rafael~C Nunez, Brittany Hagedorn,
  Lauren George, et~al.
\newblock Covasim: an agent-based model of covid-19 dynamics and interventions.
\newblock \emph{medRxiv}, 2020.

\bibitem[Balcan et~al.(2010)Balcan, Gon{\c{c}}alves, Hu, Ramasco, Colizza, and
  Vespignani]{balcan2010modeling}
Duygu Balcan, Bruno Gon{\c{c}}alves, Hao Hu, Jos{\'e}~J Ramasco, Vittoria
  Colizza, and Alessandro Vespignani.
\newblock Modeling the spatial spread of infectious diseases: The global
  epidemic and mobility computational model.
\newblock \emph{Journal of computational science}, 1\penalty0 (3):\penalty0
  132--145, 2010.

\bibitem[Dukic et~al.(2012)Dukic, Lopes, and Polson]{dukic2012tracking}
Vanja Dukic, Hedibert~F Lopes, and Nicholas~G Polson.
\newblock Tracking epidemics with state-space seir and google flu trends.
\newblock \emph{Unpublished manuscript}, 2012.

\bibitem[Osthus et~al.(2017)Osthus, Hickmann, Caragea, Higdon, and
  Del~Valle]{osthus2017forecasting}
Dave Osthus, Kyle~S Hickmann, Petru{\c{t}}a~C Caragea, Dave Higdon, and Sara~Y
  Del~Valle.
\newblock Forecasting seasonal influenza with a state-space sir model.
\newblock \emph{The annals of applied statistics}, 11\penalty0 (1):\penalty0
  202, 2017.

\bibitem[Sebastian and Victor(2017)]{sebastian2017state}
Elizabeth Sebastian and Priyanka Victor.
\newblock A state space approach for sir epidemic model.
\newblock \emph{International Journal of Difference Equations}, 12\penalty0
  (1):\penalty0 79--87, 2017.

\bibitem[Keeling et~al.(2020)Keeling, Hollingsworth, and
  Read]{keeling2020efficacy}
Matt~J Keeling, T~Deirdre Hollingsworth, and Jonathan~M Read.
\newblock Efficacy of contact tracing for the containment of the 2019 novel
  coronavirus (covid-19).
\newblock \emph{J Epidemiol Community Health}, 74\penalty0 (10):\penalty0
  861--866, 2020.

\bibitem[Sameni(2020)]{sameni2020mathematical}
Reza Sameni.
\newblock Mathematical modeling of epidemic diseases; a case study of the
  covid-19 coronavirus.
\newblock \emph{arXiv preprint arXiv:2003.11371}, 2020.

\bibitem[Godio et~al.(2020)Godio, Pace, and Vergnano]{godio2020seir}
Alberto Godio, Francesca Pace, and Andrea Vergnano.
\newblock Seir modeling of the italian epidemic of sars-cov-2 using
  computational swarm intelligence.
\newblock \emph{International Journal of Environmental Research and Public
  Health}, 17\penalty0 (10):\penalty0 3535, 2020.

\bibitem[Kobayashi et~al.(2020)Kobayashi, Sugasawa, Tamae, and
  Ozu]{kobayashi2020predicting}
Genya Kobayashi, Shonosuke Sugasawa, Hiromasa Tamae, and Takayuki Ozu.
\newblock Predicting intervention effect for covid-19 in japan: state space
  modeling approach.
\newblock \emph{BioScience Trends}, 2020.

\bibitem[Kermack and McKendrick(1927)]{kermack1927contribution}
William~Ogilvy Kermack and Anderson~G McKendrick.
\newblock A contribution to the mathematical theory of epidemics.
\newblock \emph{Proceedings of the royal society of london. Series A,
  Containing papers of a mathematical and physical character}, 115\penalty0
  (772):\penalty0 700--721, 1927.

\bibitem[Hooker et~al.(2011)Hooker, Ellner, Roditi, and
  Earn]{hooker2011parameterizing}
Giles Hooker, Stephen~P Ellner, Laura De~Vargas Roditi, and David~JD Earn.
\newblock Parameterizing state--space models for infectious disease dynamics by
  generalized profiling: measles in ontario.
\newblock \emph{Journal of The Royal Society Interface}, 8\penalty0
  (60):\penalty0 961--974, 2011.

\bibitem[Zhong et~al.(2009)Zhong, Huang, and Song]{zhong2009simulation}
ShaoBo Zhong, QuanYi Huang, and DunJiang Song.
\newblock Simulation of the spread of infectious diseases in a geographical
  environment.
\newblock \emph{Science in China Series D: Earth Sciences}, 52\penalty0
  (4):\penalty0 550--561, 2009.

\bibitem[Rapolu et~al.(2020)Rapolu, Nutakki, Rani, and Bhavani]{rapolu2020time}
Taarak Rapolu, Brahmani Nutakki, T~Sobha Rani, and S~Durga Bhavani.
\newblock A time-dependent seird model for forecasting the covid-19
  transmission dynamics.
\newblock \emph{medRxiv}, 2020.

\bibitem[Piccolomiini and Zama(2020)]{piccolomiini2020monitoring}
Elena~Loli Piccolomiini and Fabiana Zama.
\newblock Monitoring italian covid-19 spread by an adaptive seird model.
\newblock \emph{MedRxiv}, 2020.

\bibitem[Korolev(2021)]{korolev2021identification}
Ivan Korolev.
\newblock Identification and estimation of the seird epidemic model for
  covid-19.
\newblock \emph{Journal of econometrics}, 220\penalty0 (1):\penalty0 63--85,
  2021.

\bibitem[Loli~Piccolomini and Zama(2020)]{loli2020monitoring}
Elena Loli~Piccolomini and Fabiana Zama.
\newblock Monitoring italian covid-19 spread by a forced seird model.
\newblock \emph{PloS one}, 15\penalty0 (8):\penalty0 e0237417, 2020.

\bibitem[Tiwari et~al.(2020)Tiwari, Bisht, and Deyal]{tiwari2020mathematical}
Vipin Tiwari, Nandan Bisht, and Namrata Deyal.
\newblock Mathematical modelling based study and prediction of covid-19
  epidemic dissemination under the impact of lockdown in india.
\newblock \emph{medRxiv}, 2020.

\bibitem[Zipf(1946)]{zipf1946p}
George~Kingsley Zipf.
\newblock The p 1 p 2/d hypothesis: on the intercity movement of persons.
\newblock \emph{American sociological review}, 11\penalty0 (6):\penalty0
  677--686, 1946.

\bibitem[Truscott and Ferguson(2012)]{truscott2012evaluating}
James Truscott and Neil~M Ferguson.
\newblock Evaluating the adequacy of gravity models as a description of human
  mobility for epidemic modelling.
\newblock \emph{PLoS Comput Biol}, 8\penalty0 (10):\penalty0 e1002699, 2012.

\bibitem[Chen et~al.(2021)Chen, Yan, Huang, and Zhang]{chen2021correlation}
Qun Chen, Jiao Yan, Helai Huang, and Xi~Zhang.
\newblock Correlation of the epidemic spread of covid-19 and urban population
  migration in the major cities of hubei province, china.
\newblock \emph{Transportation Safety and Environment}, 3\penalty0
  (1):\penalty0 21--35, 2021.

\bibitem[Allen et~al.(2020)Allen, Altae-Tran, Briggs, Jin, McGee, Shi,
  Raghavan, Kamariza, Nova, Pereta, et~al.]{allen2020population}
William~E Allen, Han Altae-Tran, James Briggs, Xin Jin, Glen McGee, Andy Shi,
  Rumya Raghavan, Mireille Kamariza, Nicole Nova, Albert Pereta, et~al.
\newblock Population-scale longitudinal mapping of covid-19 symptoms, behaviour
  and testing.
\newblock \emph{Nature Human Behaviour}, 4\penalty0 (9):\penalty0 972--982,
  2020.

\bibitem[Buitrago-Garcia et~al.(2020)Buitrago-Garcia, Egli-Gany, Counotte,
  Hossmann, Imeri, Ipekci, Salanti, and Low]{buitrago2020occurrence}
Diana Buitrago-Garcia, Dianne Egli-Gany, Michel~J Counotte, Stefanie Hossmann,
  Hira Imeri, Aziz~Mert Ipekci, Georgia Salanti, and Nicola Low.
\newblock Occurrence and transmission potential of asymptomatic and
  presymptomatic sars-cov-2 infections: A living systematic review and
  meta-analysis.
\newblock \emph{PLoS medicine}, 17\penalty0 (9):\penalty0 e1003346, 2020.

\bibitem[Wan and Van Der~Merwe(2000)]{wan2000unscented}
Eric~A Wan and Rudolph Van Der~Merwe.
\newblock The unscented kalman filter for nonlinear estimation.
\newblock In \emph{Proceedings of the IEEE 2000 Adaptive Systems for Signal
  Processing, Communications, and Control Symposium (Cat. No. 00EX373)}, pages
  153--158. Ieee, 2000.

\bibitem[Bastos~Filho et~al.(2008)Bastos~Filho, de~Lima~Neto, Lins, Nascimento,
  and Lima]{bastos2008novel}
Carmelo~JA Bastos~Filho, Fernando~B de~Lima~Neto, Anthony~JCC Lins, Antonio~IS
  Nascimento, and Marilia~P Lima.
\newblock A novel search algorithm based on fish school behavior.
\newblock In \emph{Systems, Man and Cybernetics, 2008. SMC 2008. IEEE
  International Conference on}, pages 2646--2651. IEEE, 2008.

\bibitem[Bastos-Filho and Nascimento(2013)]{bastos2013enhanced}
CJA Bastos-Filho and DO~Nascimento.
\newblock An enhanced fish school search algorithm.
\newblock In \emph{Computational Intelligence and 11th Brazilian Congress on
  Computational Intelligence (BRICS-CCI \& CBIC), 2013 BRICS Congress on},
  pages 152--157. IEEE, 2013.

\bibitem[Tan et~al.(2020)Tan, Neto, and Braga-Neto]{tan2020pallas}
Yukun Tan, Fernando~Lima Neto, and Ulisses Braga-Neto.
\newblock Pallas: Penalized maximum likelihood and particle swarms for
  inference of gene regulatory networks from time series data.
\newblock \emph{IEEE/ACM Transactions on Computational Biology and
  Bioinformatics}, 2020.

\bibitem[Simon(2006)]{simon2006optimal}
Dan Simon.
\newblock \emph{Optimal state estimation: Kalman, H infinity, and nonlinear
  approaches}.
\newblock John Wiley \& Sons, 2006.

\bibitem[S{\"a}rkk{\"a}(2013)]{sarkka2013bayesian}
Simo S{\"a}rkk{\"a}.
\newblock \emph{Bayesian filtering and smoothing}.
\newblock Number~3. Cambridge University Press, 2013.

\bibitem[Ito and Xiong(2000)]{ito2000gaussian}
Kazufumi Ito and Kaiqi Xiong.
\newblock Gaussian filters for nonlinear filtering problems.
\newblock \emph{IEEE transactions on automatic control}, 45\penalty0
  (5):\penalty0 910--927, 2000.

\bibitem[Wu et~al.(2006)Wu, Hu, Wu, and Hu]{wu2006numerical}
Yuanxin Wu, Dewen Hu, Meiping Wu, and Xiaoping Hu.
\newblock A numerical-integration perspective on gaussian filters.
\newblock \emph{IEEE Transactions on Signal Processing}, 54\penalty0
  (8):\penalty0 2910--2921, 2006.

\bibitem[Kokkala et~al.(2015)Kokkala, Solin, and
  S{\"a}rkk{\"a}]{kokkala2015sigma}
Juho Kokkala, Arno Solin, and Simo S{\"a}rkk{\"a}.
\newblock Sigma-point filtering and smoothing based parameter estimation in
  nonlinear dynamic systems.
\newblock \emph{arXiv preprint arXiv:1504.06173}, 2015.

\bibitem[Asai(2020)]{asai2020covid}
Takashi Asai.
\newblock Covid-19: accurate interpretation of diagnostic tests—a statistical
  point of view, 2020.

\bibitem[Dong~E()]{jhu}
Gardner~L Dong~E, Du~H.
\newblock An interactive web-based dashboard to track covid-19 in real time.
\newblock \emph{Lancet Inf Dis.}, 20\penalty0 (5):\penalty0 533--534.
\newblock \doi{10.1016/S1473-3099(20)30120-1}.

\end{thebibliography}

\section*{Appendix}

\subsection*{Calculation of the covariance matrix \textbf{Q} of the process noise.}

Covariance matrix $Q_t$ is a 5 by 5 square matrix giving the covariance between each pair of elements of the state  process noise vector $(n^t_{S_i}, n^t_{E_i}, n^t_{I_i}, n^t_{R_i}, n^t_{D_i})$ for a specific region $i$ at time $t$.

Diagonal elements of the covariance matrix contain the variances of each variable, which are calculated as:

\begin{equation}
\begin{aligned}
Var (n_{S_{i}}^t ) \,=\,& Var \left(\Sigma_{j=1}^G N_{S_{i,j}}^t \right)\\[0.5ex]
\,=\,& \Sigma_j S_i^t \frac{c_{ij} \lambda_S I_j^t}{ N_j}(1 - \frac{c_{ij} \lambda_S I_j^t}{ N_j}) - \Sigma_{j_1}\Sigma_{j_2} S_i^t \frac{c_{ij_1} \lambda_S I_{j_1}^t}{ N_{j_1}} \frac{c_{ij_2} \lambda_S I_{j_2}^t}{ N_{j_2}}, \hspace{3mm} j_1 \neq j_2 \\[0.5ex]
\,=\,& \Sigma_j S_i^t \frac{c_{ij} \lambda_S I_j^t}{ N_j} - \Sigma_{j_1}\Sigma_{j_2} S_i^t \frac{c_{ij_1} \lambda_S I_{j_1}^t}{ N_{j_1}} \frac{c_{ij_2} \lambda_S I_{j_2}^t}{ N_{j_2}}
\end{aligned}
\end{equation}

\begin{equation}
\begin{aligned}
Var(n_{E_i}^t) \,=\,& Var \left(\Sigma_{j=1}^G N_{S_{i,j}}^t \right) + Var(N_{E_i}^t) \\[0.5ex]
\,=\,& \Sigma_j S_i^t \frac{c_{ij} \lambda_S I_j^t}{ N_j} - \Sigma_{j_1}\Sigma_{j_2} S_i^t \frac{c_{ij_1} \lambda_S I_{j_1}^t}{ N_{j_1}} \frac{c_{ij_2} \lambda_S I_{j_2}^t}{ N_{j_2}} + E_i^t \lambda_E(1 - \lambda_E) \\[0.5ex]
\end{aligned}
\end{equation}

\begin{equation}
\begin{aligned}
Var(n_{I_i}^t) \,=\,& Var(N_{E_i}^t) + Var(N_{R_i}^t) + Var(N_{D_i}^t) + 2Cov(N_{R_i}^t, N_{D_i}^t)\\[0.5ex]
\,=\,& E_i^t \lambda_E(1 - \lambda_E) + I_i^t \lambda_R(1 - \lambda_R) + I_i^t \lambda_D(1 - \lambda_D) - 2I_i^t\lambda_R \lambda_D \\[0.5ex]
\,=\,& E_i^t \lambda_E(1 - \lambda_E) + I_i^t(\lambda_R + \lambda_D)(1 - (\lambda_R + \lambda_D)) \\[0.5ex]
\end{aligned}
\end{equation}

\begin{equation}
Var(n_{R_i}^t) \,=\, Var(N_{R_i}^t) = I_i^t \lambda_R(1 - \lambda_R)
\end{equation}

\begin{equation}
Var(n_{D_i}^t) \,=\, Var(N_{D_i}^t) = I_i^t \lambda_D(1 - \lambda_D)
\end{equation}

The off-diagonal elements contain the covariances of each pair of variables, which are calculated as:

\begin{equation}
Cov(n_{S_i}^t, n_{E_i}^t) \,=\, - Var(n_{S_i}^t)
\end{equation}

\begin{equation}
Cov(n_{E_i}^t, n_{I_i}^t) \,=\, - Var(N_{E_i}^t)
\end{equation}

\begin{equation}
\begin{aligned}
Cov(n_{I_i}^t, n_{R_i}^t) \,=\,& - Var(N_{R_i}^t) - Cov(N_{R_i}^t, N_{D_i}^t) \\[0.5ex]
\,=\,& - I_i^t \lambda_R (1 - \lambda_R) + I_i^t \lambda_R \lambda_D \\[0.5ex]
\,=\,& - I_i^t \lambda_R (1 - (\lambda_R + \lambda_D)) \\[0.5ex]
\end{aligned}
\end{equation}

\begin{equation}
\begin{aligned}
Cov(n_{I_i}^t, n_{D_i}^t) \,=\,& - Var(N_{D_i}^t) - Cov(N_{R_i}^t, N_{D_i}^t) \\[0.5ex]
\,=\,& - I_i^t \lambda_D (1 - \lambda_D) + I_i^t \lambda_R \lambda_D \\[0.5ex]
\,=\,& - I_i^t \lambda_D (1 - (\lambda_R + \lambda_D)) \\[0.5ex]
\end{aligned}
\end{equation}

\begin{equation}
\begin{aligned}
Cov(n_{R_i}^t, n_{D_i}^t) \,=\, -I_i^t \lambda_R \lambda_D \\[0.5ex]
\end{aligned}
\end{equation}

The remaining elements will be zero.

\subsection*{Covariance of the observation noise (\textbf{R})}

Covariance matrix $R_t$ is a 2 by 2 square matrix giving the covariance between each pair of elements of the observation noise vector $({v_p}_i^t, {v_q}_i^t)$ for a specific region $i$ at time $t$.

Similarly, diagonal elements of the covariance matrix contain the variances of each variable, but all the off-diagonal elements will be zero.

\begin{equation}
    \begin{aligned}
    var({v_p}_i^t) \,=\,& \varepsilon^t_1(1 - \varepsilon^t_3)(S_i^t + R_i^t) \alpha (1 - \alpha) + \varepsilon^t_2 \varepsilon^t_3(S_i^t + R_i^t) \alpha (1 - \alpha)\\[0.5ex]
    & + \varepsilon^t_2(1 - \varepsilon^t_4) I_i^t \beta (1 - \beta) + \varepsilon^t_1(\varepsilon^t_4 I_i^t + E_i^t) \beta (1 - \beta)\\[0.5ex]
    var({v_q}_i^t) \,=\,& D_i^t \beta (1 - \beta)
    \end{aligned}
\end{equation}

\end{document}